\documentclass[10pt,journal,compsoc]{IEEEtran}

\usepackage[T1]{fontenc} 
\usepackage{amsmath,amssymb,amsfonts}
\usepackage[cmintegrals]{newtxmath}
\usepackage{bm} 

\ifCLASSOPTIONcompsoc
  \usepackage[nocompress]{cite}
\else
  \usepackage{cite}        
\fi

\usepackage{graphicx, subfig, rotating, caption, xcolor}
\usepackage{booktabs, multirow, url, lscape, xspace}
\graphicspath{{./}{results/}{figs/}}
\usepackage{orcidlink}
\usepackage{siunitx}
\usepackage{balance}

\usepackage{tikz}
\usetikzlibrary{arrows, decorations.pathmorphing, positioning, decorations.markings}
\usetikzlibrary{bayesnet}

\DeclareGraphicsExtensions{%
    .pdf,.PDF,%
    .png,.PNG,%
    .jpg,.mps,.jpeg,.jbig2,.jb2,.JPG,.JPEG,.JBIG2,.JB2}

\newcounter{example}

\newcommand{\alertbox}[2][]{%
		\begin{center}
		\begin{tikzpicture}\node[rounded corners=8pt, fill=red!50!white, inner sep=2ex]{%
			\begin{minipage}{0.94\textwidth}
			\def\title{#1}
			\ifx\title\empty
			\else
				\textbf{#1}\par
			\fi
			#2
			\end{minipage}};
		\end{tikzpicture}
		\end{center}
	}

\newcommand{\sectionref}[1]{Section~\ref{#1}}
\newcommand{\figref}[1]{Fig.~\ref{#1}}

\newcommand{\Figref}[1]{Figure~\ref{#1}}
\newcommand{\tabref}[1]{Table~\ref{#1}}
\newcommand{\equationref}[1]{equation~\eqref{#1}}

\newcommand{\biomassCh}{\ensuremath{\mathrm{CH_{1.8}O_{0.5}N_{0.2}}}}
\newcommand{\lactateCh}{\ensuremath{\mathrm{CH_3CH(OH)CO_2^-}}}
\newcommand{\sulfateCh}{\ensuremath{\mathrm{SO_4^{2-}}}}
\newcommand{\acetateCh}{\ensuremath{\mathrm{CH_3COO^-}}}
\newcommand{\bicarbonateCh}{\ensuremath{\mathrm{HCO_3^-}}}
\newcommand{\dihydrogenSulfideCh}{\ensuremath{\mathrm{H_2S}}}
\newcommand{\waterCh}{\ensuremath{\mathrm{H_2O}}}
\newcommand{\ammoniumCh}{\ensuremath{\mathrm{NH_4^+}}}
\newcommand{\protonCh}{\ensuremath{\mathrm{H^+}}}
\newcommand{\oxalateCh}{\ensuremath{\mathrm{C_2O_4^{2-}}}}
\newcommand{\oxygenCh}{\ensuremath{\mathrm{O_2}}}
\newcommand{\biomassYield}{\ensuremath{Y_{\,\mathrm{XS}}}\xspace}
\newcommand{\Cmol}{\mbox{C-mol}\xspace}

\newcommand{\biomassRV}{\ensuremath{\mathrm{X}}}
\newcommand{\lactateRV}{\ensuremath{\mathrm{L}}}
\newcommand{\acetateRV}{\ensuremath{\mathrm{A}}}
\newcommand{\sulfateRV}{\ensuremath{\mathrm{SO}}}
\newcommand{\sulfideRV}{\ensuremath{\mathrm{HS}}}
\newcommand{\bicarbonateRV}{\ensuremath{\mathrm{BC}}}
\newcommand{\oxalateRV}{\ensuremath{\mathrm{OX}}}

\newcommand{\nCells}{\ensuremath{N_C}}
\newcommand{\biomassWeight}{\ensuremath{w_{\biomassRV}}}
\newcommand{\cellWeight}{\ensuremath{m_C}}

\newcommand{\Celsius}{\si{\degreeCelsius}}

\newcommand{\NormalDistrib}{\ensuremath{\mathcal{N}}}
\newcommand{\LogNormalDistrib}{\mbox{Lognormal}}
\newcommand{\ExpDistrib}{\mbox{Exp}}

\newcommand{\etal}{et~al.}

\begin{document}

\title{Inferring Microbial Biomass~Yield and Cell~Weight
       using Probabilistic Macrochemical~Modeling}
\author{Antonio~R.~Paiva\orcid{0000-0003-0901-5922},~\IEEEmembership{Senior~Member,~IEEE,}
        Giovanni~Pilloni\orcid{0000-0002-2125-0987}%
  \IEEEcompsocitemizethanks{%
    \IEEEcompsocthanksitem A.~R.~Paiva and G.~Pilloni are with
     Corporate Strategic Research,
     ExxonMobil Research and Engineering,
     Annandale, NJ, USA.\protect\\
  }
}
\maketitle

\begin{abstract}
  Growth rates and biomass yields are key descriptors used in microbiology studies to understand how microbial species respond to changes in the environment. Of these, biomass yield estimates are typically obtained using cell counts and measurements of the feed substrate. These quantities are perturbed with measurement noise however. Perhaps most crucially, estimating biomass from cell counts, as needed to assess yields, relies on an assumed cell weight. Noise and discrepancies on these assumptions can lead to significant changes in conclusions regarding the microbes' response. This article proposes a methodology to address these challenges using probabilistic macrochemical models of microbial growth. It is shown that a model can be developed to fully use the experimental data, relax assumptions and greatly improve robustness to a priori estimates of the cell weight, and provides uncertainty estimates of key parameters. This methodology is demonstrated in the context of a specific case study and the estimation characteristics are validated in several scenarios using synthetically generated microbial growth data.
\end{abstract}


\section{Introduction}

Microbial ecology is the study of how microbes interact and respond to a variety of environmental parameters. These include the understanding of how microorganisms respond to changes in substrate concentrations, temperature, pH, salinity or pressure. The responses can then be characterized along multiple dimensions, such as growth rates and biomass yields. Moreover, these further provide evidence of energy efficiency trade-offs, activation of gene pathways, syntrophic relations and competition, etc. The understanding of these aspects and their interplay could then be used to control, or at least steer, microbial cultures to encourage growth or increased yield or, conversely, to stymie either of these.

Control of microbial cultures plays a role in a number of real-world applications. For example, in bio-remediation, microbes are necessary to break down contaminants or pollutants~\cite{bouwer1993microbial}. In those situations, microbial ecology plays a role in understanding how indigenous or introduced microorganisms can be effectively stimulated to accelerate the process or made more robust to adverse conditions~\cite{wang2019microbial}. Another application includes the prevention of microbial-induced corrosion~\cite{vigneron2018damage}. The goal in those situations is to prevent naturally occurring microorganisms from interacting with equipment or infrastructure, or from producing corrosion-inducing byproducts. Yet another application is in the control and optimization of bioreactors~\cite{jeanne2017optimization}, which are used in numerous industrial processes to produce value-added organic products as effectively as possible. Thus, improving the capabilities for studying microbial communities in natural and man-made environments is critical to advancing fundamental biology research and its implications on the many dependent applications.

Characterization of how microbial cultures respond to environmental changes is a challenging task because of the difficulty in directly measuring many of the quantities of interest. This characterization is done primarily through growth rates and biomass yields quantification from laboratory enrichment cultures. Both of these are derived measurements and their accurate estimation is of paramount importance to subsequent analysis and research. Growth rates can take advantage of repeated cell counts at different times and have better established methodologies~\cite{monod1949growth, kahm2010grofit}.
Estimation of biomass yields, defined as the ratio of overall biomass produced per amount of substrate (i.e., the ``feed'' compound) consumed, are usually harder to obtain reliably~\cite{wechselberger2013}. This is at least in part because of the indirect measurements and assumptions used to obtain the estimates~\cite{molenaar2009a}. Yet, in some circumstances, yield is as important if not more so than growth rates because it reflects how effectively the microorganisms are converting the compounds in the environment (e.g., pollutants in bio-remediation) to biomass or other byproducts. Thus, improved methods for its accurate estimation are still needed.

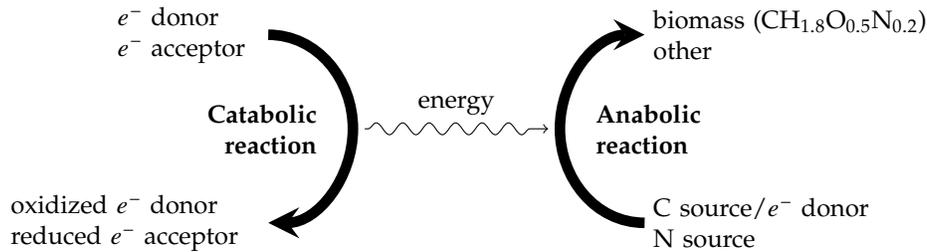
\begin{figure*}[htb]
  \begin{center}
  \begin{tikzpicture}[xscale=1.25, yscale=1.25]
    \coordinate (cat midway) at (-1.1, 0) {};
    \node[anchor=east, text width=12ex, align=right, inner sep=3ex] at (cat midway)
      {\textbf{Catabolic\\ reaction}};
    \node[anchor=east, text width=12ex] (cat inputs)  at (-2, +1)
      {$e^-$ donor\\ $e^-$ acceptor};
    \node[anchor=east, text width=21ex] (cat outputs) at (-2, -1)
      {oxidized $e^-$ donor\\ reduced $e^-$ acceptor};
    \draw[->, >=stealth, line width=4pt] (cat inputs)
      to [out=0, in=90] (cat midway)
      to [out=270, in=0] (cat outputs);
    \coordinate (anab midway) at (1.1, 0) {};
    \node[anchor=west, text width=12ex, align=left, inner sep=3ex] at (anab midway)
      {\textbf{Anabolic\\ reaction}};
    \node[anchor=west, text width=24ex] (anab inputs)  at (2, -1)  {C source/$e^-$ donor\\ N source};
    \node[anchor=west, text width=26ex] (anab outputs) at (2, +1) {biomass (\biomassCh)\\ other};
    \draw[->, >=stealth, line width=4pt] (anab inputs)
      to [out=180, in=270] (anab midway)
      to [out=90,  in=180] (anab outputs);
    \draw[->, shorten >=1ex, shorten <=1ex,
      decorate, decoration={snake, amplitude=0.5ex, pre length=2mm, post length=3mm}]
      (cat midway) -- node[above, yshift=2pt] (energy) {energy} (anab midway);
  \end{tikzpicture}
  \end{center}
  \caption{Simplified network of catabolic and anabolic metabolic processes, from Heijnen~\etal~\cite{heijnen1992a}.}
  \label{fig:macrochemical.model:process.reactions}
\end{figure*}

The characterization of microbial cultures can be enhanced by explicit modeling of microbial growth. This modeling can be done at different levels of detail and complexity. One modeling approach describes the overall growth process by a set of simplified metabolic networks and the corresponding macrochemical reactions~\cite{heijnen1992a, heijnen2010a}. The main advantages of this modeling approach are that it ensures overall stoichiometric balance requiring only knowledge of the main growth modes of the microorganisms. This model can also be used toward a thermodynamic characterization of growth energy trade-offs because it describes the amount of energy retained in the system as biomass~\cite{russell1995a, smith2007a}.

Much more involved approaches have been proposed by explicitly modeling the metabolic network of microbes. Approaches based on biochemical flow modeling or flux balance analysis~(FBA) can be used to characterize the stoichiometric balance of a number of metabolites involved in the growth process and their interaction through time~\cite{orth2010a, kauffman2003a, lee2006a}. These are typically developed for specific microbes with widely studied metabolism, such as \textit{Escherichia coli}, or microbe types under specific conditions, such as sulfate-reducing bacteria~\cite{noguera1998unified, smith2019mathematical}. FBA models are also often used in bioengineering applications to optimize yield and other growth aspects~\cite{kauffman2003a, lee2006a}. 
These models leverage significant advances in the understanding of genetic regulation of pathways and their related metabolic reactions, allowing for the simulation of intracellular metabolism dynamics, based on actual measurements of the fluxomics. Such models are understandably much more descriptive and, under an appropriate characterization of the metabolic processes and with sufficient flow measurement data, provide a more complete characterization of growth. On the other hand, these approaches must characterize a number of the metabolic reactions and their dynamics to avoid ill-posedness, which depend on the estimation of a number of kinetic parameters or flow constraints~\cite{orth2010a, noguera1998unified}. To calibrate these parameters, these approaches require repeated measurement of concentrations throughout the duration of the experiments, which can be onerous or difficult to obtain. Hence, the methodology proposed herein aims to provide more effective models when flow measurements are hard to perform or a detailed metabolic characterization is unavailable.

For these reasons, the macrochemical model approach is used as the base stoichiometric characterization in this article. In contrast to FBA and flow models, this modeling approach requires only the ability to specify the general macrochemical framework for the growth and a few general assumptions about biomass composition. Nonetheless, the probabilistic methodology by which we extend the macrochemical characterization could in principle be similarly extended using other base models, including dynamic models (e.g., FBA), albeit potentially at a significant increase in complexity and data requirements.

This article demonstrates how a macrochemical characterization for a given microbe and growth conditions can be formulated as a probabilistic graphical model. This can be done from any macrochemical characterization, as Sections~\ref{sec:srb} and \ref{sec:oxalaticus} exemplify. A key advantage of the resulting probabilistic models is that one is able to effectively reconcile all of the experimentally measured data and enforce chemical mass balances in a principled manner. As a result, it is shown in \sectionref{sec:results} that one is able to more accurately estimate underlying culture growth parameters, such as biomass yield. Perhaps most crucially, using this formulation, one can relax certain assumptions which would otherwise severely impair the analysis (see \sectionref{sec:cell.weight.changes}).

\section{Macrochemical characterization}
\label{sec:macrochem.model}

\subsection{Catabolic and anabolic processes}

The methodology described in this article is based on a simplified set of metabolic processes proposed by Heijnen and colleagues~\cite{heijnen1992a, heijnen2010a}, of which a brief overview is provided. In their framework, microbial growth is described by two macrochemical processes, corresponding to catabolic and anabolic reactions, as depicted in \figref{fig:macrochemical.model:process.reactions}. We refer to these as macrochemical processes as they characterize the overall inputs and outputs of these in aggregate.
Of these two processes, the catabolic reaction characterizes the main energy generating process, wherein the substrate is oxidized to generate ATP which is then used to support the anabolic reactions of cell growth, maintenance, and biomass production.

The chemical equation for the catabolic reaction has the following general form:
\begin{multline}
  \theta_1(\mbox{$e^-$ donor}) + \theta_2(\mbox{$e^-$ acceptor}) \\
  + \theta_3(\mbox{oxidized $e^-$ donor}) \\
  + \theta_4(\mbox{reduced $e^-$ acceptor}) = 0
  \label{eq:cat.process}
\end{multline}
where $\{\theta_i: i=1,\ldots,4\}$ are the stoichiometric coefficients of the compounds involved. The electron donor and acceptor are reaction inputs (i.e., `reactants') and their oxidized and reduced counterparts are outputs (i.e., `products'). We use the convention that compounds with negative coefficients correspond to consumed reactants and positive coefficients to products. The coefficient of the electron donor is usually set to $\theta_1=-1$ to avoid an arbitrary scaling indeterminacy, and then the other coefficients are determined accordingly.

Depending on the microbe and the mode of respiration, this can be an aerobic or anaerobic process. In aerobic growth, oxygen is used as the electron acceptor, whereas in anaerobic growth another reactant (e.g., sulfate) is reduced. Although we focus on the anaerobic case here, both cases are exemplified in \sectionref{sec:pgm}.

The other key metabolic process is described by the anabolic reaction, which characterizes the production of biomass. For heterotrophic growth the anabolic chemical equation has the general form:
\begin{multline}
  \alpha_1(\mbox{C source/$e^-$ donor}) + \alpha_2(\mbox{N source}) + \alpha_3\mbox{H$^+$} \\
  + \biomassCh + \alpha_4(\mbox{oxidized $e^-$ donor}) + \alpha_5\waterCh = 0
  \label{eq:anab.process}
\end{multline}
with $\{\alpha_i\}$ the corresponding stoichiometric coefficients, determined such that biomass coefficient was equal to one.

The generic compound formula \biomassCh{} is used to denote biomass, representing the aggregate relative proportion of elements in dry cell biomass~\cite{heijnen1992a}. This composition ignores other elements, such as S, P, K, and Mg, because of their small fraction ($< 1-2\%$)~\cite{heijnen2010a}. For details on the experimental details leading to this composition the reader is referred to Battley~\cite{battley1998development}. An analysis of the effect of variations in this assumed biomass composition is shown in the supplementary material~S1. The biomass amount is measured in `carbon moles' (\Cmol units), corresponding to the amount of ash-free organic biomass containing one mole of carbon~\cite{heijnen2010a, battley1998development}, and the other elements in their respective relative proportions according to the assumed biomass compound formula (cf.~\sectionref{sec:biomass.est}).

The catabolic and anabolic processes cannot be directly combined because there is an unknown and condition dependent scaling factor between the two. This unknown catabolic scaling factor, also known as ``metabolic quotient''~\cite{righelato1968a} or ``catabolic turnover rate'', is inversely proportional to how efficiently the microbe can use energy generated via the catabolic reaction toward biomass production. From a macrochemical balance perspective, it tell us how many times the catabolic reaction needs to run for each mole of biomass produced. 
Since the primary goal here is the estimation of biomass yield and this can be done directly, we do not infer the scaling factor. If this was a quantity of interest, however, it could be estimated from the biomass yield. The reader is referred to Heijnen and Kleerebezem~\cite{heijnen2010a} for details and examples.


\subsection{Biomass estimation}\label{sec:biomass.est}

Estimating the amount of biomass produced is crucial for characterizing the anabolic process. This can be done through a number of methods~\cite{madrid2005microbial}. A well-known method is by direct measurement of dry cell organic weight. This is a simple but very laborious and time consuming method~\cite{battley1998development}. A much simpler method involves first obtaining cell counts through microscopy or flow cytometry. Biomass can then be estimated using relative proportions of elements in dry cell biomass and average cell weight. With regard to the above-mentioned biomass compound formula, \biomassCh{}, the molar mass of the elements and their relative proportion can be combined to obtain the biomass molar mass as $\biomassWeight = \SI{24.62}{\gram\per\Cmol}$~\cite{heijnen1992a}.

Assuming a given average cell weight \cellWeight, corresponding to the organic biomass dry weight in grams of a cell, biomass can then be estimated as
\begin{equation}
  \biomassRV = (\nCells \times \cellWeight) / \biomassWeight
  \label{eq:biomass}
\end{equation}
where $\nCells$ denotes the cell count.

\subsection{Biomass yield}\label{sec:biomass.yield}

The biomass yield can now be clearly established. It is defined as,
\begin{equation}
  \biomassYield
    = \frac{\mbox{C-moles of biomass produced}}{\mbox{moles of $e^-$ donor consumed}}
  \label{eq:defn.biomass.yield}
\end{equation}
As highlighted by the definition, the biomass yield is notable because it indicates how efficiently the consumption of electron donor (i.e., substrate) is used toward the production of biomass. This is a crucial element in the characterization of how a microbial culture responds to changes in environmental conditions, such as temperature and substrate concentrations.

Without noise, the estimation of biomass yield would be straightforward from the definition and experimental measurements. In practice, however, measurement noise and experimental variability can make it quite difficult to discern the underlying trend or profile in the response of a microbial culture. Moreover, using cell counts, biomass estimation relies on assumptions on cell weight, which may not hold. By appropriately characterizing both the catabolic and anabolic processes using \emph{all of the available data}, the estimation of biomass produced and electron donor consumed can be improved significantly and made much more robust to deviations from assumptions.


\section{Probabilistic macrochemical modeling}
\label{sec:pgm}

This section now describes how a macrochemical model of growth can be formulated into a probabilistic model. Specifically, this model will be defined using the formalism of Bayesian graphical models~\cite{koller2009probabilistic, gelman2014bayesian}. This is a general methodology that might have to be adapted for a different microbe or for different growth conditions because of changes to the corresponding catabolic and anabolic chemical equations (equations~\eqref{eq:cat_equation} and \eqref{eq:anab_equation}). For this reason, this section emphasizes the process used in mapping the macrochemical equations to the probabilistic model formulation used for statistical inference afterward.

In order to concretely demonstrate how the probabilistic macrochemical model methodology can be applied, this process will be demonstrated with respect to two different bacteria grown under different conditions. The first example case study will consider a sulfate-reducing bacteria~(SRB) grown under anaerobic conditions, and the second case considers bacterial aerobic growth in oxalate. The first case will be presented in full detail to completely walk the reader through the completely process from macrochemical equations to code implementation. This will also be the microbial growth case considered in the validation experiments in \sectionref{sec:results}. For the second aerobic case, we will show only the crucial changes to the reformulation, from which one could easily modify the implementation for the first case.

\subsection{SRB growth case study}\label{sec:srb}

\subsubsection{Macrochemical equations}\label{sec:srb.macrochem.eqs}

Consider a situation in which sulfate-reducing bacteria~(SRB) are grown with lactate (\lactateCh) under anaerobic conditions~\cite{muyzer2008ecology}. In this case, one can easily verify that lactate will serve both the role of electron donor and carbon source, and thus plays a role in both the catabolic and anabolic processes. The additional necessary compounds for growth can also be easily identified to include sulfate (\sulfateCh), and ammonium (\ammoniumCh) as the nitrogen source. Accordingly, the balanced catabolic and anabolic chemical equations are, respectively,
\begin{gather}
  \begin{split}
    &-\lactateCh - 0.5\sulfateCh \\
    &\quad\qquad + \acetateCh + \bicarbonateCh + 0.5\dihydrogenSulfideCh = 0
    \label{eq:cat_equation}
  \end{split}\\
  \begin{split}
    &-0.35\lactateCh - 0.2\ammoniumCh - 0.1\protonCh \\
    &\qquad + \biomassCh + 0.05\bicarbonateCh + 0.4\waterCh = 0
    \label{eq:anab_equation}
  \end{split}
\end{gather}
which show acetate (\acetateCh), bicarbonate (\bicarbonateCh), and (dihydrogen) sulfide (\dihydrogenSulfideCh) as byproducts. The stoichiometric coefficients were determined to ensure the balance of mass, charge, and degree of reduction of each equation.


\subsubsection{Model specification}\label{sec:srb.pgm}

Bayesian models, and our probabilistic macrochemical model in particular, are \emph{generative} explanations of the experimental data. This means that the model describes our understanding of how the growth interdependencies between reactants and products lead to the recorded measurements. The macrochemical reactions completely reflect our understanding of the growth processes and chemical/mass balance constraints that must be asserted. Then, the probabilistic model translates the macrochemical reactions between the true unknown value of the quantities of interest and the noisy measurements collected.

With respect to equations~\eqref{eq:cat_equation} and \eqref{eq:anab_equation}, we will assume that the observed (i.e., measured) data comprises cell counts and concentration measurements for lactate, sulfate, acetate, bicarbonate and sulfide. The measurements are obtained at the beginning and end of the experiments, such that they characterize the steady-state change in each of the compounds. We consider that the data pertains to several conditions between which we which to assess differences in growth yield, of which a single condition is obviously a special case. However, for simplicity, it will be considered that the experiments have similar initial starting amounts of compounds and cells and thus we focus on modeling the change in the different compounds and cells. If this were not the case, the different initial conditions and the dependencies on these could be easily accounted for in the model as necessary to explain the experimental circumstances. Moreover, if cell counts or concentrations measurements for either of these compounds are not available, then the corresponding equations that would characterize those measurements can simply be left out from the model. Understandably, this leads to a less constrained model and therefore likely to have larger uncertainty about the inference results. In fact, this scenario is tested and shown in the results section when analyzing model results under different situations.

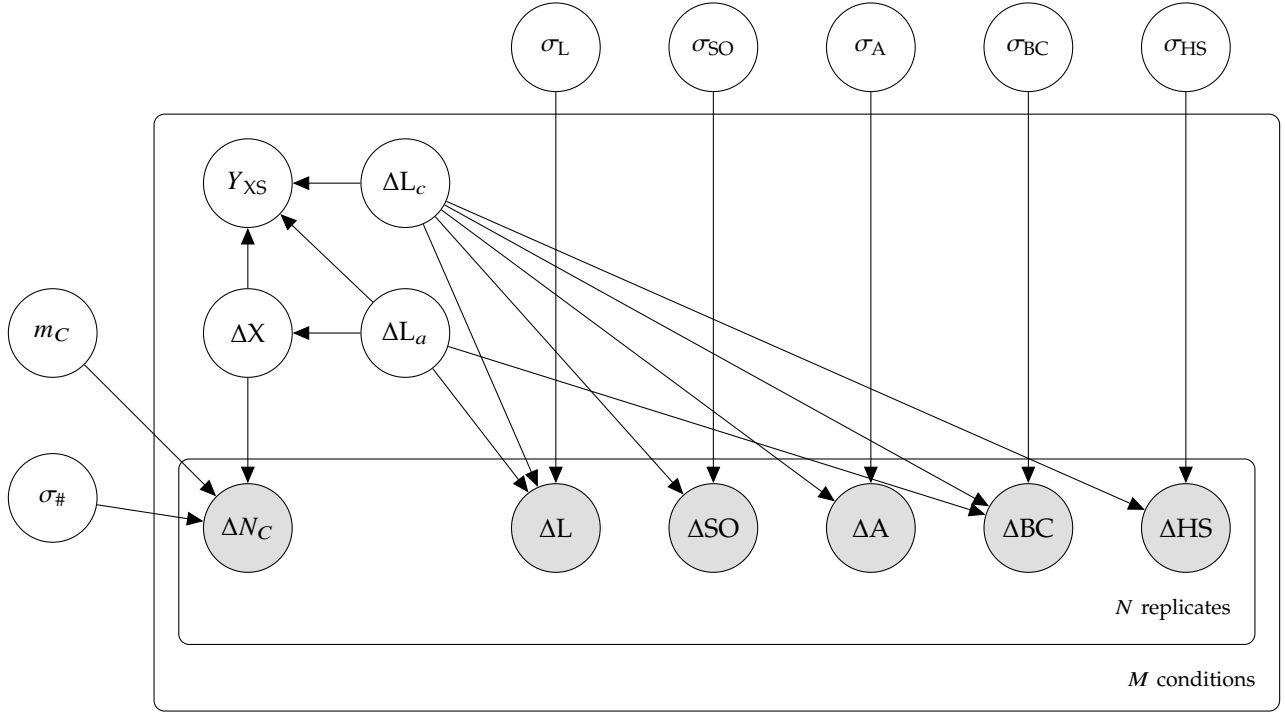
\begin{figure*}
  \begin{center}
    \scalebox{1.0}{%
    \begin{tikzpicture}[x=9mm, y=10mm, scale=1]
      \tikzstyle{latent} += [minimum width=7.5ex]
      \tikzstyle{plate} += [inner sep=2ex]
      \tikzstyle{plate caption} += [yshift=-1ex]

      \node[latent]                   (dLcat)       {$\Delta\lactateRV_c$};
      \node[latent, below=of dLcat, yshift=2mm]   (dLanab)      {$\Delta\lactateRV_a$};
      \node[latent, left=of dLcat, xshift=0mm] (bYield)  {$\biomassYield$};
      \node[obs, below=of dLanab, yshift=-4mm, xshift=20mm] (dL) {$\Delta\lactateRV$};
      \node[obs, right=of dL]         (dSO)         {$\Delta\sulfateRV$};
      \node[obs, right=of dSO]        (dA)          {$\Delta\acetateRV$};
      \node[obs, right=of dA]         (dBC)         {$\Delta\bicarbonateRV$};
      \node[obs, right=of dBC]        (dHS)         {$\Delta\sulfideRV$};
      \node[latent, left=of dLanab, xshift=0mm] (dX)        {$\Delta\biomassRV$};
      \node[latent, left=of dX, xshift=-5mm]    (mCell)     {$\cellWeight$};
      \node[obs, below=of dX, yshift=-4mm]      (nCells)    {$\Delta\nCells$};
      \node[latent, below=of mCell, yshift=0mm] (sNcells)   {$\sigma_{\#}$};
      \node[latent, above=of dL,  yshift=42mm] (sL)  {$\sigma_{\lactateRV}$};
      \node[latent, right=of sL]  (sSO) {$\sigma_{\sulfateRV}$};
      \node[latent, right=of sSO] (sA)  {$\sigma_{\acetateRV}$};
      \node[latent, right=of sA]  (sBC) {$\sigma_{\bicarbonateRV}$};
      \node[latent, right=of sBC] (sHS) {$\sigma_{\sulfideRV}$};

      \edge {dLanab, dLcat, sL} {dL};
      \edge {dLanab, dLcat, sBC} {dBC};
      \edge {dLcat, sSO} {dSO};
      \edge {dLcat, sA}  {dA};
      \edge {dLcat, sHS} {dHS};
      \edge {dLanab, dLcat, dX} {bYield};
      \edge {dLanab} {dX};
      \edge {dX, mCell, sNcells} {nCells};

      \plate {forEachReplicate}
        {(nCells) (dL) (dBC) (dSO) (dA) (dHS)}
        {$N$ replicates};
      \plate {forEachCondition}
        {(dLanab) (dLcat) (dX) (bYield) 
         (forEachReplicate.north west) (forEachReplicate.south east)}
        {$M$ conditions} ;
    \end{tikzpicture}%
    }
  \end{center}
  \caption{Plate notation diagram of the probabilistic macrochemical model. Each node in the diagram denotes a random variable, with shaded ones being observed random variables. In other words, we have noisy measurements of the random variables in the shaded nodes, whereas the other nodes are completely defined by the model.
    The \lactateRV, \sulfateRV, \acetateRV, \bicarbonateRV, and \sulfideRV qualifiers, used either with respect to concentration changes (e.g., $\Delta\lactateRV$) or $\sigma$ measurement noise variables, indicate the lactate, sulfate, acetate, bicarbonate, or sulfide compound that they refer to. $\Delta\biomassRV$ denotes the amount of biomass produced and $\Delta\nCells$ the increase in the total number of cells as a result, which depend of the cell weight $\cellWeight$ and cell count error $\sigma_{\#}$. $\biomassYield$ denotes the target estimate of biomass yield.
    The ``plates'' (i.e., boxes enclosing multiple nodes) indicate that the random variables within are indexed for each of the number of situations noted on the lower right corner of the plate. Hence, the probabilistic model infers a different distribution for each of the situations.}
  \label{fig:pgm}
\end{figure*}

There are a number of ways to reformulate the macrochemical reactions into the probabilistic macrochemical model but, since lactate is a key compound and consumed in both reactions, it provides a good way to connect the two metabolic processes. Accordingly, we define two latent random variables corresponding to the amount of lactate consumed in the catabolic reaction, $\Delta\lactateRV_c$, and in the anabolic reaction, $\Delta\lactateRV_a$. These quantities are unknown but, with respect to them, one can directly specify the expected changes to the other compounds and in either of the processes by stoichiometric balance.
Another key random variable is the amount of biomass produced which is related to the lactate used in anabolic reaction by mass balance and the cell count by \equationref{eq:biomass}. These three random variables are central to the generative nature of the model in the sense that, if they were known, they are crucial to explaining the observed measurements except for noise/experimental variability. A few other unknown parameters (e.g., cell weight) and auxiliary random variables are used to complete the specification.

With regard to the two latent quantities of lactate consumed, the probabilistic model characterization of the change in each of the measured quantities can be written as,
\begin{align}
  \mbox{Lactate:}\ &
    \Delta\lactateRV \sim \NormalDistrib(
      \Delta\lactateRV_c + \Delta\lactateRV_a, 2\sigma_\lactateRV^2)
      \label{eq:delta.total.lactate}\\
  \mbox{Sulfate:}\ &
    \Delta\sulfateRV \sim \NormalDistrib(
      0.5\Delta\lactateRV_c, 2\sigma_{\sulfateRV}^2)
      \label{eq:delta.total.sulfate}\\
  \mbox{Acetate:}\ &
    \Delta\acetateRV \sim \NormalDistrib(
      -\Delta\lactateRV_c, 2\sigma_\acetateRV^2) \\
  \mbox{Bicarbonate:}\ &
    \Delta\bicarbonateRV \sim \NormalDistrib(
      -\Delta\lactateRV_c - \frac{0.05}{0.35}\Delta\lactateRV_a,
      2\sigma_{\bicarbonateRV}^2) \\
  \mbox{Sulfide:}\ &
    \Delta\sulfideRV \sim \NormalDistrib(
      -0.5\Delta\lactateRV_c, 2\sigma_{\sulfideRV}^2)
      \label{eq:delta.total.sulfide} \\
  \mbox{Cell counts:}\ &
    \Delta\nCells \sim \NormalDistrib(
      (\biomassWeight\Delta\biomassRV)/\cellWeight, \sigma_{\#}^2)
      \label{eq:delta.cell.count}
\end{align}
where the notation $\sim \NormalDistrib(\mu, \sigma^2)$ indicates that a given quantity is modeled as being normal distributed with mean $\mu$ and variance $\sigma^2$, $\biomassWeight=\SI{24.64}{g/\Cmol}$ is the biomass molar mass, and $\cellWeight$ denotes the average cell weight. The various $\sigma^2$ variance hyperparameter random variables reflect the unknown concentration measurement error of each of the corresponding compounds. Although these could be estimated directly through extensive experimental replication, the approach shown here will prove quite data efficient. Since the change in each of the compounds is obtained from the difference of the initial and final measurements, the measurement error is introduced twice. Hence the $2\times$ factor in the variances of concentration measurements. With regard to the cell counts, we considered the count variability to be relative to the cell density and thus the variability in the initial cell counts to be negligible. Accordingly, $\sigma_{\#}^2$ is determined by the final cell count in the normal distribution, and needs to be accounted for only once.

The probabilistic model equations can be easily interpreted with respect to the formulation and stoichiometry in the macrochemical equations. The first equation, \equationref{eq:delta.total.lactate}, simply states that the amount of lactate consumed between the catabolic and anabolic reactions must be explainable by the measured change of lactate in the substrate, except for small differences due to measurement noise (with variance $\sigma_\lactateRV^2$). The other equations generatively explain the remaining observed random variables (i.e., with measurements). For example, \equationref{eq:delta.total.sulfate} reflects the observation from the catabolic equation that only 0.5 moles of sulfate are consumed for every mole of lactate. Then, this probabilistic macrochemical model is to infer the distributions of the unknown random variables that best explains the observed data.

This probabilistic model is represented in \figref{fig:pgm}. The diagram highlights the dependencies between the two catabolic and anabolic lactate consumed, and the biomass produced, measurements of the different chemical compounds, and their estimated measurement noise level. The arrows in the diagram denote a model factorization in which the distribution of random variables on destination nodes is conditional on the value of the starting nodes. Note the plates aggregating the different replicate measurements (i.e., repetition of the experiment under the same conditions to account for inherent variability) and conditions (e.g., experiments at different temperature or pressure conditions). Intuitively speaking, the goal is to combine multiple samples whenever possible such as to robustly estimate quantities across conditions, while accurately characterizing condition specific quantities.

Using the above characterization of lactate consumed in each metabolic process, the biomass produced and yield can be directly obtained as
\begin{align}
  \Delta\biomassRV  &= \frac{1}{-0.35}\Delta\lactateRV_a \\
  \biomassYield     &= \frac{\Delta\biomassRV}
                            {-(\Delta\lactateRV_c + \Delta\lactateRV_a)}
                    = \frac{\Delta\lactateRV_a}
                           {0.35(\Delta\lactateRV_c + \Delta\lactateRV_a)}
                      \label{eq:yield:srb}
\end{align}
Actually, the inference process yields a distribution for each of these due to Markov Chain Monte Carlo sampling~(MCMC) algorithm used~\cite{gelman2014bayesian}.

Note that estimating the biomass produced and yield in this way is more robust than direct mass balance between measurements because all of the available data is simultaneously accounted for by the model, even if cell counts are not available. And, if cell counts can be used, these estimates can be further validated with regard to the observed counts. Using cell counts, the biomass yield could in principle have been calculated externally to the probabilistic model. However, including it herein is much more robust and provides us with uncertainty on the estimates which can be helpful diagnostic information.

It must also be emphasized that the average cell weight, $\cellWeight$, is shown in \figref{fig:pgm} as common to all experiments. This assumption can be relaxed, which can be key to identifying cell weight changes due to environmental stresses in the experiments. This actual scenarios will considered and validated in the results (\sectionref{sec:cell.weight.changes}).

Finally, the prior distributions of the random variables need to be defined to complete the model specification. They were set to:
\begin{align}
  \Delta\lactateRV_c &\sim \NormalDistrib(-28, 0.5^2) \\
  \Delta\lactateRV_a &\sim \NormalDistrib(-2, 0.5^2) \\
  \cellWeight &\sim \LogNormalDistrib(\ln(\num{1.8e-13}) + 0.09, 0.3)
\end{align}
\begin{align}
  \sigma_{\#} &\sim \ExpDistrib(1/10^8) \label{eq:sigma.nCells:prior}\\
  \sigma_{\lactateRV} &\sim \ExpDistrib(1/\varsigma) \\
  \sigma_{\sulfateRV} &\sim \ExpDistrib(1/\varsigma) \\
  \sigma_{\acetateRV} &\sim \ExpDistrib(1/\varsigma) \\
  \sigma_{\bicarbonateRV} &\sim \ExpDistrib(1/\varsigma) \\
  \sigma_{\sulfideRV} &\sim \ExpDistrib(1/\varsigma)
\end{align}
with $\varsigma=0.1$. These priors were chosen assuming an initial lactate concentration in solution of \SI{30}{mM}, as is considered in the validation section. The lactate is usually almost fully consumed during growth. The prior cell weight was set such that the mode is at \SI{1.8e-13}{\gram}, corresponding to the average cell weight of \textit{E.~coli} reported in Fagerbakke\etal~\cite{fagerbakke1996}. The priors on cell counts and concentration measurement standard deviations were set to exponential distributions (with means $10^8$ and $\varsigma$, respectively) to enforce their positivity and discourage unwarranted attribution of mismatch between estimates to measurement variability. The mean of the cell count distribution was chosen to be about an order of magnitude smaller than the final cell counts. The analysis is not very sensitive to the parameters for the $\sigma$ priors provided that they assign sufficient density to plausible values.

\subsubsection{Implementation and inference}

The above equations completely specify the probabilistic model, which needs to be implemented such as to allow for statistical inference from data. In other words, to estimate the posterior distributions given the data. Since the model is non-parametric, inference needs to be done using a Markov Chain Monte Carlo~(MCMC) sampling algorithm, such as Hamiltonian Monte Carlo~(HMC)~\cite{gelman2014bayesian}.
In our case, this particular model was implemented using the Stan probabilistic programming language~\cite{carpenter2017stan, gelman2014bayesian}. In Stan, the code specifies a generative model of the data, which is entirely specified by the above equations as can be observed in the implementation included in the supplementary material~S2. In fact, one can observe that the Stan model specification code reflects the above equations nearly line-by-line.

From this code, the Stan compiler produces an executable for performing inference on the model latent parameters when applied to data. The compiled model can efficiently perform inference of all model quantities in just a few seconds. The complete code used to interface with the Stan model and obtain the validation results is available at \url{github.com/arpaiva/biopgm-macrochem}.

\subsection{\textit{Pseudomonas oxalaticus} aerobic growth example}\label{sec:oxalaticus}

As previously mentioned, probabilistic macrochemical modeling is a general methodology that can be adapted for any microbe under a variety of growth conditions given a characterization via catabolic and anabolic chemical equations. Hence, to contrast with the previous case study, we now exemplify the key steps of the formulation for aerobic growth of \textit{Pseudomonas oxalaticus} on oxalate (\oxalateCh)~\cite{heijnen1992a}. As before, oxalate will be involved in both the catabolic and anabolic processes as electron donor and carbon source, respectively. It should be emphasized that this dual role of the primary growth compound, oxalate in this case, is not in any way a requirement but that is often the case for heterotrophs and facilitates the modeling somewhat by explicitly introducing a dependency between quantities from both equations and a measured random variable. Of course, for aerobic growth the catabolic equation will also require oxygen (\oxygenCh). Ammonium (\ammoniumCh) is used as the nitrogen source in the anabolic reaction as before.
Accordingly, the balanced catabolic and anabolic chemical equations are, respectively,
\begin{gather}
  \begin{split}
    &-2\oxalateCh - 2\waterCh - \oxygenCh + 4\bicarbonateCh = 0
    \label{eq:cat_equation:oxalate}
  \end{split}\\
  \begin{split}
    &-0.5\oxalateCh - 0.2\ammoniumCh - 0.1\waterCh - 0.8\protonCh \\
    &\qquad\qquad + \biomassCh + 0.8\oxygenCh = 0
    \label{eq:anab_equation:oxalate}
  \end{split}
\end{gather}
which show bicarbonate (\bicarbonateCh) as the main byproduct of the catabolic reaction. Again, the stoichiometric coefficients were determined to ensure the balance of mass, charge, and degree of reduction in each equation.

Then, consider that in this cases the experimental data comprises cell counts and concentration measurements for oxalate and bicarbonate, obtained at the beginning and end of the growth experiments. If the experiments have similar initial concentrations and cell counts, then we can focus on modeling the difference in conditions at the end of each experiment as a result of growth processes. Although there are several ways to formulate the model, since oxalate is used in both the catabolic and anabolic processes, it can be used to connect connect the random variables from the two equations. As before, we define two latent random variables, $\Delta\oxalateRV_c$ and $\Delta\oxalateRV_a$, corresponding to the amount of oxalate consumed in the catabolic and anabolic reactions, respectively. With respect to these two latent quantities of oxalate consumed, the probabilistic macrochemical model characterization of the change in the other compounds can be written as,
\begin{align}
  \mbox{Oxalate}\ &
    \Delta\oxalateRV \sim \NormalDistrib(
      \Delta\oxalateRV_c + \Delta\oxalateRV_a, 2\sigma_\oxalateRV^2)
      \label{eq:delta.total.oxalate}\\
  \mbox{Bicarbonate:}\ &
    \Delta\bicarbonateRV \sim \NormalDistrib(
      -2\Delta\oxalateRV_c, 2\sigma_{\bicarbonateRV}^2) \\
  \mbox{Cell counts:}\ &
    \Delta\nCells \sim \NormalDistrib(
      (\biomassWeight\Delta\biomassRV)/\cellWeight, \sigma_{\#}^2)
      \label{eq:delta.cell.count:oxalate}
\end{align}
where $\sigma_\oxalateRV^2$ is the variance random variable for the oxalate concentrate measure error. The other random variables and biomass molar mass $\biomassWeight=\SI{24.64}{g/\Cmol}$ are denoted as before. We also assumed the same considerations pertaining to the variance of cell counts. The probabilistic models is completed by the specification of the priors on the random variables as previously demonstrated.

And, from this characterization the biomass produced and yield in this case can be readily estimated as,
\begin{align}
  \Delta\biomassRV &= \frac{1}{-0.5}\Delta\oxalateRV_a \\
  \biomassYield    &= \frac{\Delta\biomassRV}{
                             -(\Delta\oxalateRV_c + \Delta\oxalateRV_a)}
                    = \frac{2\Delta\oxalateRV_a}
                           {\Delta\oxalateRV_c + \Delta\oxalateRV_a}
\end{align}

Compared to the previous case study, this example results in a simpler overall probabilistic macrochemical model because fewer compounds are measured. The fewer degrees of freedom are likely to make the MCMC inference algorithm converge faster. On the other hand, this also means that there is less data and stoichiometry interdependencies constraining the model, likely resulting in higher uncertainty (i.e., broader distribution) of the posterior estimate of the random variables. As illustrated in \figref{fig:chem.model2:yield}(d), higher uncertainty can always be countered by collecting additional data.

Given the simplicity of the macrochemical characterization, one might question the added benefit of the probabilistic model formulation in this example. If the goal is to analyze a single condition with a sufficient number of replicates, that may well be the case. However, the main advantages to estimation using a probabilistic macrochemical model should be apparent in tackling more complex experimental questions.
For example, if there are multiple conditions with potential confounding factors (e.g., cell weight \cellWeight) to the estimation of quantities of interest, as considered in \sectionref{sec:cell.weight.changes}, then the probabilistic model enables us to perform rigorous hypothesis testing. Or if there are additional measurements leading to different avenues to estimate the quantity of interest, as considered in the previous case study, then the probabilistic model enables us to simultaneously consider all of the data and to weigh each measurement according to their jointly inferred measurement noise.

\section{Validation results}\label{sec:results}

We now consider two simulated microbial growth scenarios of the SRB case study for validation of the corresponding probabilistic macrochemical model and its inference estimates. Although the derived model could have been shown here using real-data, we resort to simulated scenarios so that the model and inference results can be quantitatively compared against the underlying, ground-truth parameters values used in the simulation.

The results of the probabilistic macrochemical model will be compared to common direct estimation approaches that a microbiologist would use to estimate these quantities. As stated in the introduction, we focus here on situations where only initial and final concentration measurements are available. Hence, this work tackles experimental circumstances for which flow-based modeling approaches, even in the specific context of SRBs (e.g., \cite{noguera1998unified, smith2019mathematical}), are not applicable because the lack of repeated concentration measurements during the experiments. When that data is available, those models are generally better suited as the repeated intermediate measurements act as multipliers on the number of replicates, whereas models based on the macrochemical characterization considered herein are unable to leverage such information.


Most crucially, it must be emphasized that the primary goal of this section is to demonstrate the benefits of translating a simple macrochemical characterization of the microbe's growth into a probabilistic macrochemical model with which to analyze the experimental measurements.

\subsection{Testing paradigm}
\label{sec:data.gen}

\begin{table*}
  \caption{Generated data of change in concentrations (in \si{mM}) and cell counts for both scenarios considered for validation. Note that only the cell counts change between scenarios (cf.~\sectionref{sec:cell.weight.changes}).}
  \label{tab:data.combined}
  \begin{center}
    \begin{tabular}{rccccccc}
  \toprule
  Temp (\Celsius) & $\Delta\lactateRV$ & $\Delta\sulfateRV$
    & $\Delta\acetateRV$ & $\Delta\bicarbonateRV$ & $\Delta\sulfideRV$
    & $\Delta\nCells$ (Scenario~1) & $\Delta\nCells$ (Scenario~2) \\
  \midrule
     5 &  -29.89 &  -12.71 &  27.02 &  25.73 &  13.52 &  3.20e+09 &  1.98e+09 \\
     5 &  -29.86 &  -13.64 &  27.21 &  28.86 &  14.51 &  3.44e+09 &  2.22e+09 \\
     5 &  -30.05 &  -13.46 &  27.22 &  28.20 &  13.54 &  3.37e+09 &  2.15e+09 \\
    10 &  -30.30 &  -12.92 &  27.69 &  28.32 &  12.91 &  3.58e+09 &  2.80e+09 \\
    10 &  -29.82 &  -13.48 &  27.33 &  27.06 &  12.62 &  3.08e+09 &  2.31e+09 \\
    10 &  -30.27 &  -13.56 &  27.10 &  28.70 &  13.74 &  3.29e+09 &  2.51e+09 \\
    15 &  -30.19 &  -13.17 &  27.02 &  25.88 &  15.04 &  3.39e+09 &  3.15e+09 \\
    15 &  -29.62 &  -13.75 &  26.91 &  26.11 &  14.07 &  3.15e+09 &  2.92e+09 \\
    15 &  -30.27 &  -13.67 &  27.63 &  28.96 &  14.27 &  2.98e+09 &  2.74e+09 \\
    20 &  -29.66 &  -14.61 &  28.26 &  30.23 &  13.59 &  1.91e+09 &  1.89e+09 \\
    20 &  -30.58 &  -13.96 &  28.18 &  28.89 &  14.34 &  2.21e+09 &  2.19e+09 \\
    20 &  -30.13 &  -15.04 &  28.52 &  27.76 &  13.95 &  2.16e+09 &  2.15e+09 \\
    25 &  -29.52 &  -14.92 &  29.31 &  30.70 &  14.63 &  6.01e+08 &  5.71e+08 \\
    25 &  -29.94 &  -14.88 &  29.68 &  29.75 &  13.53 &  5.37e+08 &  5.08e+08 \\
    25 &  -29.87 &  -14.92 &  29.61 &  28.97 &  14.94 &  4.86e+08 &  4.56e+08 \\
  \bottomrule
\end{tabular}

  \end{center}
\end{table*}

To generate the data for testing the model, we consider an hypothetical set of experiments in which one would like to recover a microorganism's biomass yield profile with regard to temperature.
The simulated scenarios comprise $M=5$ temperature conditions ($T = \{5, 10, 15, 20, 25\}\Celsius$) and $N=3$ replicate experiments per temperature condition. The underlying (i.e., ground-truth) biomass yield curve is shown as a dashed line in \figref{fig:model.free:yield} and similar subsequent figures. This particular yield curve was chosen such as to mimic the biomass yield values and profile of strain LSv21 (\textit{Desulfofrigus fragile}), a psychrophilic SRB bacteria with a temperature optimum of 18\Celsius, studied in Knoblauch and J{\o}rgensen~\cite{knoblauch1999effect}, and for which the macrochemical equations in \sectionref{sec:srb.macrochem.eqs} apply. The paper from Knoblauch and J{\o}rgensen~\cite{knoblauch1999effect} investigates the optimal growth temperature, assessed through rates of sulfate reduction and growth yields, of psychrophilic SRB isolated from permanently cold marine sediment. Out of 5 strains analyzed in that work, we focused on LSv21, affiliated to \textit{Desulfofrigus fragile}, which has optimum growth at 18\Celsius but it is still able to grow at 0\Celsius, albeit at 31\% of the growth rate at optimum. 

For each temperature condition, the data is obtained by essentially applying the macrochemical metabolic processes in equations~\eqref{eq:cat_equation} and \eqref{eq:anab_equation} for the underlying biomass yield curve. All experiments were considered to start from the same (ground truth) initial concentrations of the measured chemical compounds and cell counts:
\begin{align*}
  \mbox{Lactate:}\ & \SI{30}{mM} \\
  \mbox{Sulfate:}\ & \SI{30}{mM} \\
  \mbox{Acetate:}\ & \SI{0.1}{mM} \\
  \mbox{Bicarbonate:}\ & \SI{20}{mM} \\
  \mbox{Sulfide:}\ & \SI{0.2}{mM} \\
  \mbox{Cell count:}\ & 10^7
\end{align*}
Then, using the biomass yield response for each condition (i.e., temperature in this case), one can determine the total amount of biomass produced, considering that the lactate was almost fully consumed except for a small residual amount. Note however that, in spite of the common starting point, the final concentrations of the compounds differ significantly between conditions since the yield is dependent on the condition temperature. Then, the biomass yield curve establishes for each condition the split of the amount of lactate consumed in each of the catabolic and anabolic reactions. Accordingly, the amount consumed or produced of each of the other compounds can then be sampled according to equations~\eqref{eq:delta.total.lactate} through \eqref{eq:delta.cell.count}. Similarly, the observed change in concentrations and cell counts can be obtained. This calculation is deterministic and establishes the \emph{ground-truth values} for each condition.

In practice, one must deal with measurement noise and inherent experimental variability. Therefore, all measurements were obtained by sampling according to equations~\eqref{eq:delta.total.lactate} through \eqref{eq:delta.cell.count} with the standard deviations shown in \tabref{tab:noise.sigmas}.
Those standard deviations were chosen to mimic the levels of measurement noise seen in actual experimental measurements and differences in accuracy between different concentration measurement techniques~\cite{hallbeck2014determination}.
Note that this amount of noise would be applied to both initial and final concentration measurements equations, hence the $2\times$ factor on variances on the observed change in concentration values from equations~\eqref{eq:delta.total.lactate}--\eqref{eq:delta.total.sulfide}. For cell counts, since the amount of noise is likely proportional to the value, the noise on the initial count is assumed negligible and applied only to the final value.
The generated data is shown in \tabref{tab:data.combined}.

\subsection{Scenario~1: Constant cell weight scenario}
\label{sec:const.cell.weight}


We consider first a scenario in which the cell weight was constant across conditions. Understandably, this simplifies the analysis and is well suited to the assumptions commonly used by practitioners.

\begin{figure}
  \begin{center}
    \subfloat[Estimates from concentration measurements of
      lactate and the listed compound.]{
      \includegraphics[width=0.46\textwidth
        ]{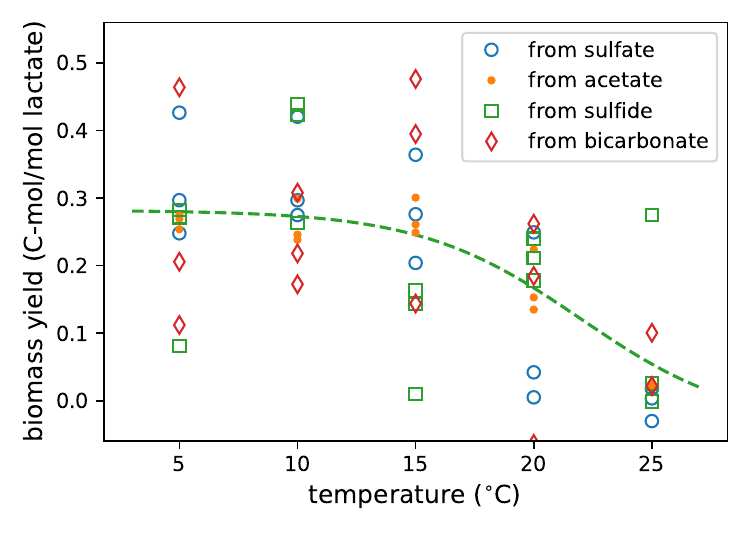}}
    \\
    \subfloat[Estimates from lactate concentration and cell counts.]{
      \includegraphics[width=0.46\textwidth
        ]{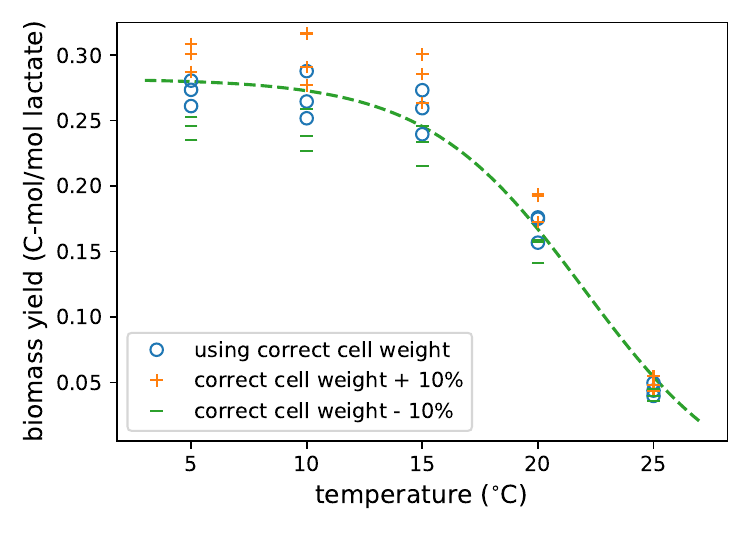}}
  \end{center}
  \caption{Biomass yield estimates for Scenario~1 using only direct concentration measurements in (a) and by including cell counts in (b). The estimates are obtained by establishing mass balance between the catabolic and anabolic reactions; please refer to \sectionref{sec:const.cell.weight} for details.
    For reference, the underlying trend curve used in generating the data is indicated by the dashed line.}
  \label{fig:model.free:yield}
\end{figure}

As a baseline, and for the sake of argument, it is worth considering to what extent the biomass yield profile could be estimated using only concentration measurements of two compounds and their stoichiometric balance. For example, from \equationref{eq:cat_equation} and the measurement of how much sulfate was consumed, one can estimate the amount of lactate consumed in the catabolic reaction as twice that. Thus, the amount of lactate consumed in the anabolic reaction can be estimated as the total lactate consumed minus that used in the catabolic reaction. Substituting those estimates in \equationref{eq:yield:srb} yields the biomass yield. The results using the total lactate consumed and separately for each of the other measured compounds are shown in \figref{fig:model.free:yield}(a). While hints of the underlying trend are visible, it is quite difficult to outline it because of the large variability. Additionally, there is significant variability in those results depending on which compound concentrations are used because of their different measurement noise, which adds uncertainty regarding the most likely trend.

As previously mentioned, a common practice is to first estimate biomass produced using cell counts according to \equationref{eq:biomass} and then use that results in calculating the yield. These results are shown in \figref{fig:model.free:yield}(b). In this example, since a constant and known cell weight was used in generating the data, the yield estimates are quite accurate. However, as shown in the figure, if the assumed cell weight is off, the yield estimates will be offset accordingly.

\begin{figure}
  \begin{center}
    \subfloat[Estimates from concentration measurements only but combined via the probabilistic macrochemical model.]{
      \includegraphics[width=0.46\textwidth
        ]{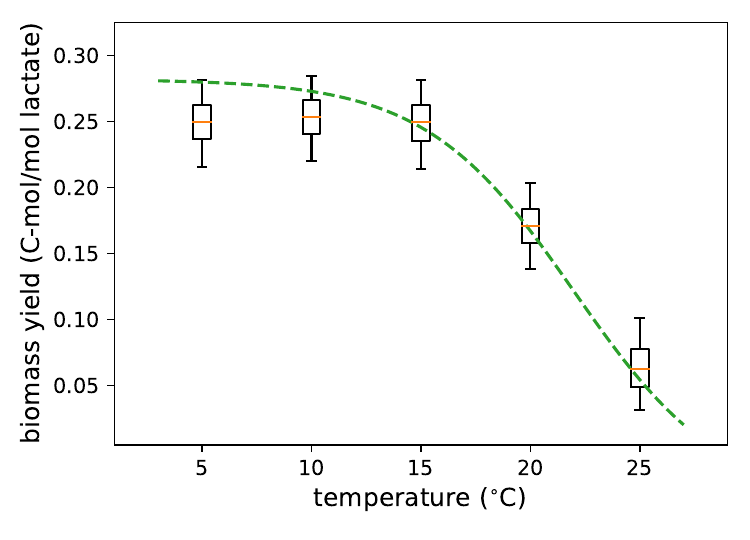}}
    \\
    \subfloat[Estimates from all concentration measurements and cell counts.]{
      \includegraphics[width=0.46\textwidth
        ]{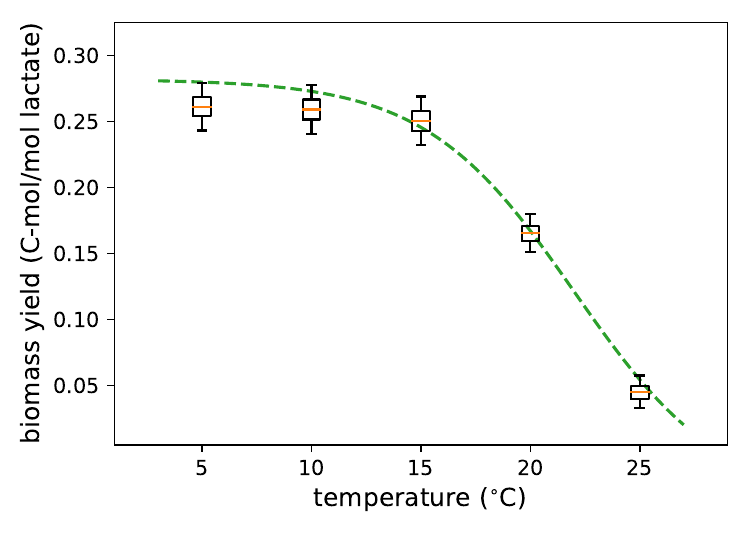}}
  \end{center}
  \caption{Biomass yield estimates using the probabilistic macrochemical model for Scenario~1. In line with the scenario's conditions, the cell weight is assumed to be the same for all conditions. Compare these plots with those in \figref{fig:model.free:yield}.}
  \label{fig:chem.model1:yield}
\end{figure}

Those results can be contrasted to the biomass mass yield estimates shown in \figref{fig:chem.model1:yield} using the probabilistic macrochemical model developed in \sectionref{sec:pgm}. Even if the cell counts and the corresponding latent variables are not used, one can observe from \figref{fig:chem.model1:yield}(a) that the yield estimates already reflect most of the underlying trend. As expected and clearly shown in \figref{fig:chem.model1:yield}(b), including the cell counts improves the accuracy of the estimates and reduces their uncertainty. It is worth emphasizing that the estimated trends of yield decreasing with increasing temperature (rather than peaking at 18\Celsius) match those observed by Knoblauch and J{\o}rgensen~\cite{knoblauch1999effect}. Still, arguably one of the most significant advantages of the proposed approach is that, since the cell weight is estimated by the probabilistic model, the results are robust to large variations in the initial cell weight, as illustrated in \figref{fig:chem.model1:yield}. Those results are not shown as there was no discernible difference in the biomass yield estimation results, even when the cell weight used to generate the data was changed by as much as $\pm 50\%$ of the probabilistic model prior value.

\begin{table}
  \caption{Noise standard deviations. The values used to generate the data are shown on the left, and the estimates in Scenario~1 using the probabilistic macrochemical model on the right. Concentration standard deviations are in \si{mM}.}
  \label{tab:noise.sigmas}
  \begin{center}
    \begin{tabular}{ccc}
      \toprule
      Noise & Ground truth value & Mean estimated value\\ \midrule
      $\sigma_{\lactateRV}$     & 0.15 & 0.20 \\
      $\sigma_{\sulfateRV}$     & 0.3  & 0.30 \\
      $\sigma_{\acetateRV}$     & 0.2  & 0.18 \\
      $\sigma_{\bicarbonateRV}$ & 0.8  & 0.71 \\
      $\sigma_{\sulfideRV}$     & 0.4  & 0.45 \\
      $\sigma_{\#}$             & \num{2e8} & \num{1.74e8} \\
      \bottomrule
    \end{tabular}
  \end{center}
\end{table}

In addition, our probabilistic model also infers the distributions for the noise standard deviations. As previously mentioned, those model variables aim to explain measurement noise and experimental variability. The mean estimates of the noise standard deviations are shown in Table~\ref{tab:noise.sigmas}.
Even with only 15 experiments (5 conditions $\times$ 3 replicates), the mean estimates of each of the concentration standard deviations (i.e., the $\sigma$'s) were within 12\% of the value used in generating the data, except for lactate. The mean of $\sigma_{\lactateRV}$ was $36\%$ larger, likely because of its central role in the model, meaning that the inferred distribution for $\sigma_{\lactateRV}$ played a significant role in explaining other measurements and not only its own. The estimated cell count standard deviation was also within 13\% of the ground-truth value.
In any event, it is worth reiterating that these estimates are not a goal per se. Rather, their accurate estimation is noteworthy because it demonstrates the ability of the probabilistic macrochemical model in optimally combining all the measurements accounting for the potential noise variance in each.

\subsection{Scenario~2: Cell weight changes with temperature}
\label{sec:cell.weight.changes}


\begin{figure}
 \begin{center}
   \includegraphics[width=0.46\textwidth]{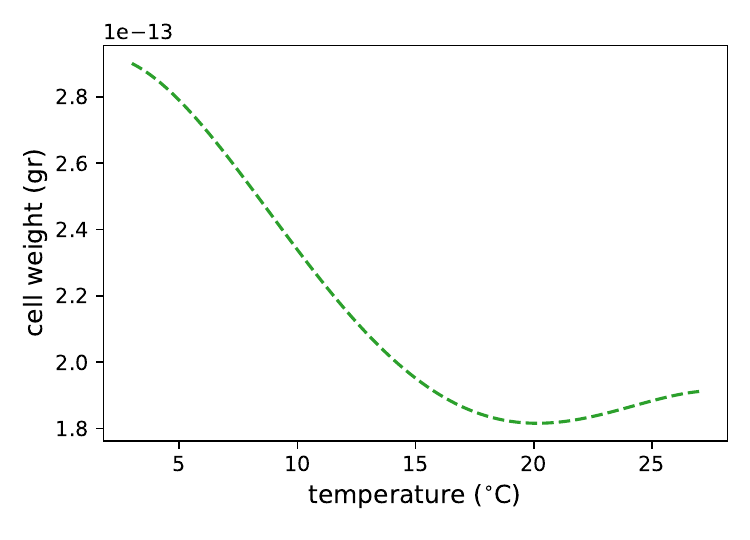}
 \end{center}
 \caption{Cell weight temperature response curve used to generate the data for Scenario~2 in \sectionref{sec:cell.weight.changes}. Although the biomass production is as in Scenario~1 (as this is controlled by the yield curve), this changes the number of the cells which contain that biomass. In other words, this changes the amount of biomass per cell.}
 \label{fig:cell.weight.profile}
\end{figure}

Assuming that the cell weight is unaffected by the experimental conditions can be a significant source of analysis error. Several studies have reported changes in cell weight or volume in response to different environmental conditions (see, for example, Pavlovsky~\etal~\cite{pavlovsky2015effects} or Wiebe~\etal~\cite{wiebe1992bacterial}).
Hence, it is worth analyzing how the proposed methodology might perform in such a situation, how it compares to standard practices, and how it may be extended if needed.

The data used for this analysis was obtained using the same procedure as before, detailed in \sectionref{sec:data.gen}, but in which the cell weight $\cellWeight$ for each temperature condition was determined according to \figref{fig:cell.weight.profile}.
The values of the cell weight curve were chosen by scaling by \SI{1.8e-13}{\gram\per cell} according to the proportions reported in Wiebe~\etal~\cite[Table~1]{wiebe1992bacterial} which studied the effect of nutrient concentration on growth rates and cell biovolumes at low temperatures on bacteria isolated from marine samples.
It is important to emphasize that the cell weight temperature effect does not change the quantity of biomass produced, which is obtained by stoichiometric balance with respect to the biomass yield curve. Rather, it changes only how the amount of biomass produced affects the cell counts at different temperatures.


\begin{figure*}
  \begin{center}
    \subfloat[Estimates using lactate concentration measurements and cell counts directly, as shown in \figref{fig:model.free:yield}(b) for Scenario~1.]{
      \includegraphics[width=0.46\textwidth]{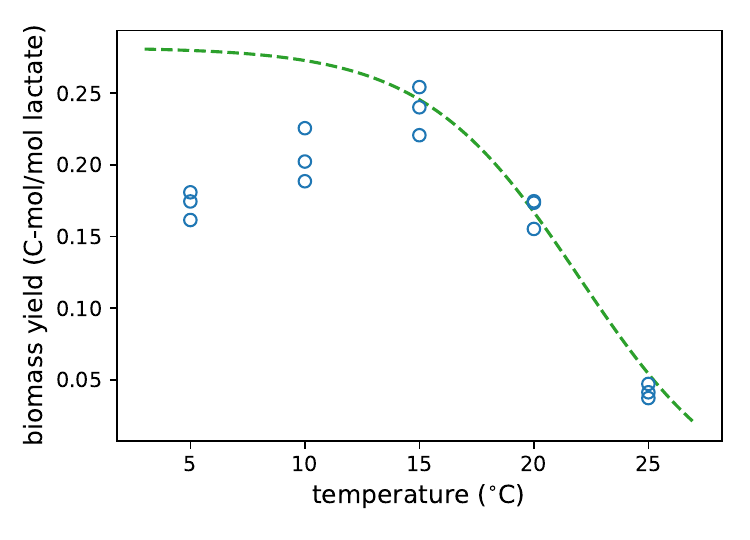}}
    \quad
    \subfloat[Estimates assuming the same cell weight for all conditions.]{
      \includegraphics[width=0.46\textwidth]{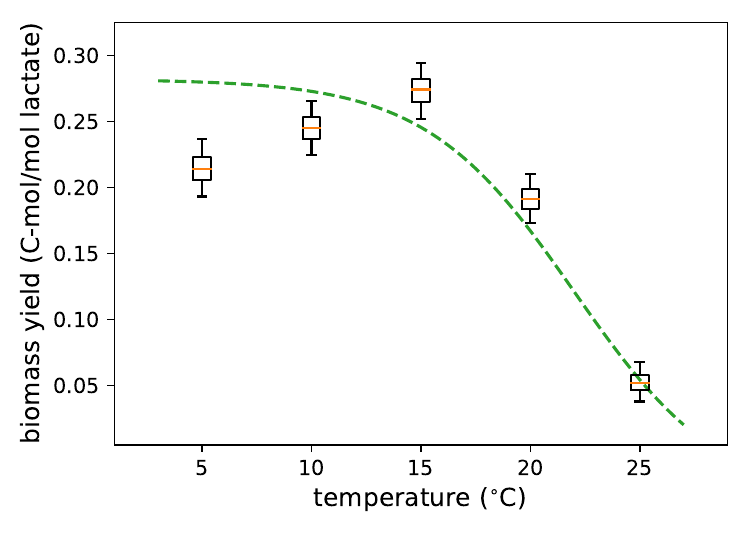}}
    \\
    \subfloat[Estimates with cell weight estimated per condition from \emph{3~replicates} per condition.]{
      \includegraphics[width=0.46\textwidth]{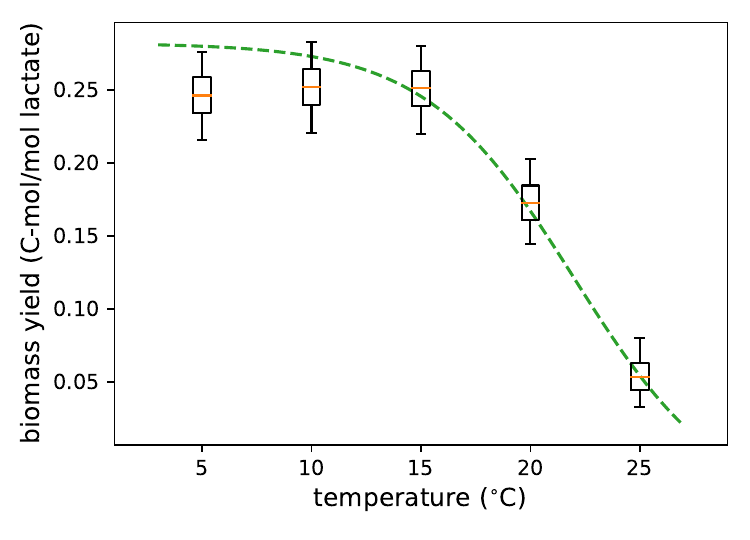}}
    \quad
    \subfloat[Estimates with cell weight estimated per condition from \emph{7~replicates} per condition.]{
      \includegraphics[width=0.46\textwidth]{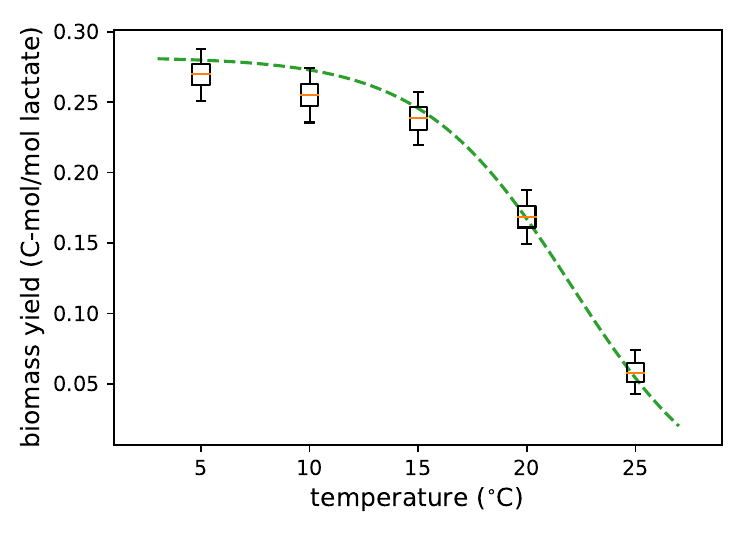}}
  \end{center}
  \caption{Biomass yield estimates for Scenario~2. (a) and (b) clearly show the systematic error in the estimates from direct mass balance estimates and the probabilistic macrochemical model with a common cell weight random variable. Changing the probabilistic model to infer the cell weight separately for each condition substantially improves the accuracy of the estimates with respect to those in (a) and (b). This is shown in (c) and (d) using a different number of replicates per condition. Unsurprisingly, using a higher number of replicates results in more accurate and yet lower uncertainty yield estimates.}
  \label{fig:chem.model2:yield}
\end{figure*}

Unsurprisingly, as shown in Figs.~\ref{fig:chem.model2:yield}(a) and \ref{fig:chem.model2:yield}(b), the assumption that the cell weight does not change in response to the conditions clearly results in biased estimates of the biomass yield profile both using direct mass balance estimation techniques and the probabilistic model \emph{with that assumption}. This is most noticeably for lower temperatures because of the larger difference in cell weights. For higher temperatures, the lower amount of biomass produced means that the effect is also proportionally lower. It is worth noting that this situation is fundamentally different than the variations in cell weight studied in the previous scenario because the assumption of constant cell weight across conditions still held true in that case.

An interesting observation is that those estimated biomass yield profiles seem to exhibit a distinct peak at \SI{18}{\celsius}, which would perhaps seem more intuitive to biologists. However, we know by design that that is not true in this case since the biomass yield curve used to generate the data is monotonically decreasing with temperature in the range considered. This is yet another example highlighting the inherent pitfalls that may arise from analyzing one's data using flawed assumptions. Nevertheless, as we will demonstrate, the uncertainty quantification provided by the probabilistic model can be instrumental in identifying and diagnosing such situations.

The probabilistic model considered thus far also infers the cell weight from data but assumes that it is the same across conditions. For that reason, it is unable to correct for this systematic error in its model assumptions (\figref{fig:chem.model2:yield}(a)). However, unlike other approaches, our derived probabilistic model also characterizes the uncertainty in the estimates, of which the posterior distribution over cell weights is highly informative in this case. In the previous scenario, in which the data aligned with the model assumptions, the standard deviation on the posterior distribution of cell weight was \SI{7.7e-15}{\gram}, whereas in this scenario is \SI{1.1e-14}{\gram}. A large uncertainty of a quantity taken to be the same for all conditions (i.e., across all data points) could be an indication of an incorrect assumption in the model. From a modeling perspective, this would support an alternate hypothesis that the environmental conditions also have an effect in the microorganism's weight, which happens to be the case in this scenario.


Testing the hypothesis of condition dependent cell weights can be done simply by modifying the probabilistic model such as to make the latent variable $\cellWeight$ a condition-specific quantity. This would be depicted in \figref{fig:pgm} by moving the $\cellWeight$ latent variable inside the conditions plate, meaning that that quantity is assumed common to experiments under the same condition (i.e., replicates) but inferred independently for each condition. The corresponding results are shown in \figref{fig:chem.model2:yield}(c). Much of the bias due to the different cell weights has been compensated for by the model. There is a larger uncertainty in the biomass yield estimates however. This was expected because the change introduces additional model parameters and additional ways to explain the observed data.

Another fundamental advantage of the probabilistic methodology is that extending the model allows us to ``explain away'' potential sources of systematic error given enough observed data. Although the added degrees of freedom of the model may result in large uncertainty in the inferred parameters, as can be observed in \figref{fig:chem.model2:yield}(c), that can be compensated by additional data. This is demonstrated in \figref{fig:chem.model2:yield}(d) in which four additional replicates ($N=7$) were used to show the improved statistical robustness while accounting for the effect of the conditions on cell weight. Current practices and models without this extension would maintain the systematic yield estimation error regardless.

\begin{figure}
  \begin{center}
    \includegraphics[width=0.46\textwidth]{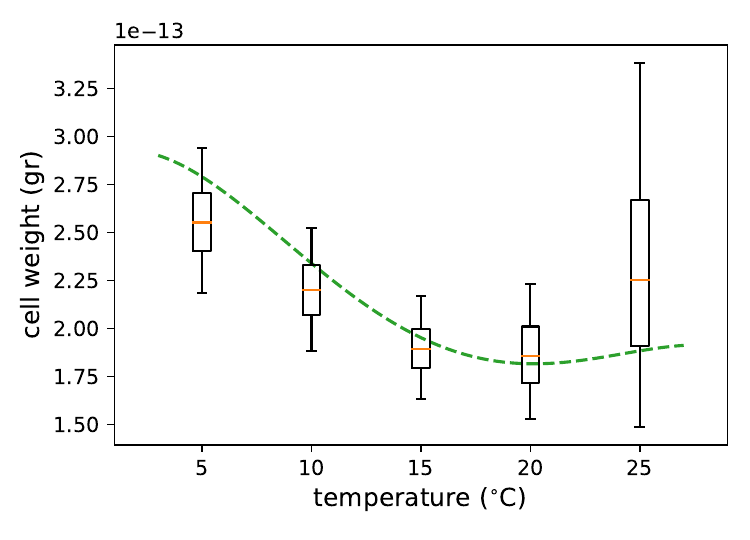}
  \end{center}
  \caption{Cell weight estimates using the probabilistic macrochemical model with $\cellWeight$ inferred per condition. The dashed line indicates the cell weight profile curve used in generating the data.}
  \label{fig:chem.model2:cell.weight}
\end{figure}

With the extended probabilistic model one also obtains estimates of the cell weights for each temperature condition. \Figref{fig:chem.model2:cell.weight} shows these results compared them to the underlying profile curve used in generating the data. The model results recover the general trend although there is substantial uncertainty because only 3 replicates per condition were used and there are other factors involved in explaining the data variability. The much larger uncertainty at $25\Celsius$ is because the cell counts are significantly smaller ($\sim 4\times$ compared to the other conditions) and thus much more susceptible to cell count noise.

\section{Discussion}

This article proposed a probabilistic methodology for characterization of microbial growth. The methodology combines the formalism of Bayesian graphical models with a macrochemical representation of cell metabolic processes. To the best of our knowledge, such a general probabilistic modeling approach has not been applied to microbial growth before. This approach has several advantages compared to the current practices. First and foremost, it allows for multiple measurements to be incorporated from first principles and used toward improving the estimation accuracy of the quantities of interest. Another key advantage is the improved robustness of the estimates under a variety of circumstances and certain deviations of model parameters. A third advantage is the characterization of uncertainty it provides via the posterior distributions over the inferred variables. This can be used to assess how well the model explains the observed data, but also to identify aspects that were difficult to fit the model to. These may suggest alternate hypothesis, which can be easily tested by extending the model given the appropriate and sufficient data or that may be candidates for subsequent experimental validation. Or, conversely, those could indicate conditions for which additional replicates should obtained to reduce the uncertainty.

While the methodology is very general, it must be expressed with respect to the particular circumstances of one microorganism and growth condition under consideration. This imposes an up-front modeling burden but, as we hope to have clearly demonstrated, this is clearly compensated for with the improved accuracy and robustness of the resulting estimates from data.

A potential criticism with regard to the testing as presented could be that the data generation mirrored several aspects of the model. While that would be understandable, it is worth emphasizing that the data generation approach also played directly to the assumptions of current practices and experiments with similar microbes performed in our laboratory. The exception was the validation scenario with different cell weights which aimed to contrast the rigidity of current practices with the flexibility of the probabilistic model to infer that experimental factor. In sum, the crucial difference is that the probabilistic macrochemical modeling methodology proposed herein provides us with approaches to effectively utilize all the data available, assess model quality via its uncertainty estimates, and easily adapt to deviations on model assumptions when necessary.

With regard to Scenario~2, it could be said that the situation and rationale to extend the model is somewhat circuitous given the scenario. In reality, the scenario was motivated by insights from the practical application of the proposed methodology to real-data. Moreover, several examples of different cell morphologies, and therefore weights, for microbial lab-cultures under various conditions have been reported in the literature~\cite{westfall2017bacterial, pavlovsky2015effects, wiebe1992bacterial} and recently also by our group~\cite{jones2020evidence}.

There are several considerations for future work. Recall that the methodology as presented characterizes conditions independently, except for common parameters that are constant through all experiments. Hence, it would be beneficial to augment the methodology such that the trend across conditions is explicitly modeled and inferred for one or more latent variables of interest. A possibility could be to use a Gaussian process prior over conditions which could encourage smooth trends while non-parametrically inferring the trend from data. This would also enable obtaining direct predictions of quantities at conditions not yet tested. Another avenue for future work is the extending and testing of this modeling methodology to cultures of multiple microorganisms.


\section*{Acknowledgements}
The authors would like to thank Frederick von~Netzer, Drew Gorman-Lewis, and David~A.~Stahl at the University of Washington for very helpful discussions.

\balance
\bibliographystyle{IEEEtran}
\bibliography{refs}

\newpage
\balance
\begin{IEEEbiography}[{\includegraphics[width=1in,height=1.25in,clip,keepaspectratio]{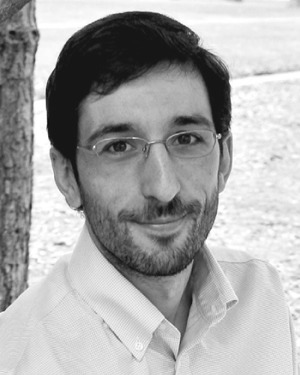}}]{Antonio R.\ Paiva} (S'05-M'09-SM'19) received the \textit{licentiate} degree in electronics and telecommunications engineering from the University of Aveiro, Portugal in 2003, and the M.S.\ and Ph.D.\ degrees in electrical and computer engineering from the University of Florida in 2005 and 2008, respectively.

Between 2003 and 2004 he was as a research assistant at the Institute of Electronics and Telematics Engineering at the University of Aveiro working on image compression. And, from 2008 to 2010, he worked at the SCI Institute at the University of Utah as a post-doctoral fellow on the application of machine learning to image analysis and neural circuit reconstruction. Between 2010 and 2015, he was a research specialist with ExxonMobil Upstream Research Company working on analysis of geophysical data. He is currently a research associate at Corporate Strategic Research in the ExxonMobil Research and Engineering Company. He is an Associate Editor of the IEEE Transactions on Neural Networks and Learning Systems, and a past Associate Editor of the IEEE Signal Processing Letters.

His research interests span several areas within machine learning, pattern recognition, and adaptive signal processing and their applications. He has published numerous articles on kernel methods, information-theoretic learning, and fast algorithms for machine learning. Most recently, his research is focused on applications of probabilistic graphical models and probabilistic deep learning models, and approximate methods for large-scale Bayesian inference.
\end{IEEEbiography}%

\begin{IEEEbiography}[{\includegraphics[width=1in,height=1.25in,clip,keepaspectratio]{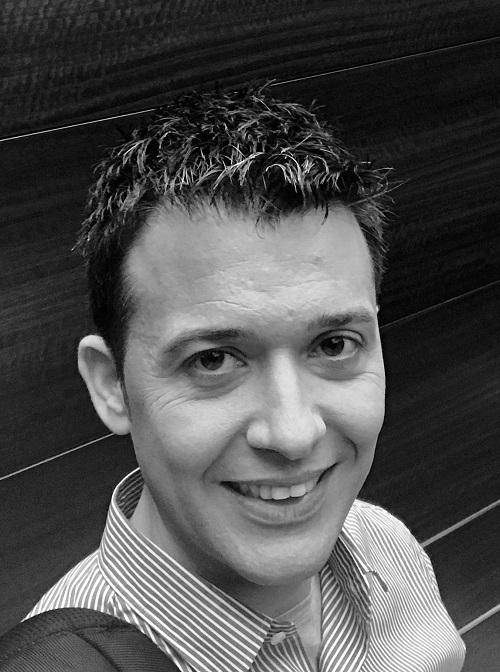}}]{Giovanni Pilloni} obtained a MS in Biology from Catania University (Italy) and a PhD in Natural Sciences from the Technical University of Munich (Germany). After a Postdoc at the Helmholtz Center of Munich, he joined the Corporate Strategic Research Lab of ExxonMobil Research and Engineering in 2012. Giovanni applied his background in Microbial Ecology to numerous fields, including freshwater ecology, bioremediation of groundwater, microbial corrosion, reservoir souring, and cellulosic biofuels, as captured by several of his contributed publications.
\vspace*{1.4in}
\end{IEEEbiography}

\end{document}


\title{Supplementary Material for\\ \huge
       ``Inferring Microbial Biomass~Yield and Cell~Weight using Probabilistic Macrochemical~Modeling''}
\author{
  Antonio~R.~Paiva and Giovanni~Pilloni\\
  Corporate Strategic Research,
  ExxonMobil Research and Engineering,
  Annandale, NJ, USA}
\maketitle

\section{Effect of biomass composition}
\newcommand{\biomassChAlt}{\ensuremath{\mathrm{CH_{1.613}O_{0.557}N_{0.158}}}}

The generic biomass composition formula $\biomassCh$ was used throughout the article~\cite{heijnen1992a, heijnen2010a}. In this section, we consider the implications of differences in the assumed biomass composition formula used in the probabilistic macrochemical model. To study this effect, we use the same data (generated with the biomass composition as before) but assume a different biomass composition formula within our probabilistic macrochemical model. This enables us to also demonstrate how to balance these equations in general.

In \cite{battley1998development}, Battley reported on experiments with \textit{Saccharomyces cerevisiae} (yeast cells) and measured the relative elemental composition from dry weight after lyophilization. Accordingly, in that study the biomass composition formula was estimated to be $\biomassChAlt$. As before, we ignore the ash components (containing P, S, K, Mg, and Ca) included in \cite{battley1998development} because of their minor contribution ($< 1-2\%$). Note that, compared to the biomass composition formula used in generating the data, this involves substantial deviations of $\approx 11\%$ in both hydrogen and oxygen and $21\%$ in nitrogen.

Considering this biomass composition in the anabolic equation of our SRB case study example will require recalculating the stoichiometric coefficients (i.e., the $\alpha$'s in equation~(2)). For anaerobic growth in lactate the anabolic equation takes the general form,
\begin{equation}
  \alpha_1\lactateCh + \alpha_2\ammoniumCh + \alpha_3\protonCh
    + \biomassChAlt + \alpha_4\bicarbonateCh + \alpha_5\waterCh = 0.
\end{equation}
Then, one can write a system of equations in which each equation balances each of the involved elements (C, H, O, and N) and electric charge (see Example~1 in Heijnen~\etal~\cite{heijnen2010a} for further details). Solving this system of equations for this case results in
\begin{equation}
  -0.3354\lactateCh - 0.158\ammoniumCh - 0.1712\protonCh
    + \biomassChAlt + 0.0063\bicarbonateCh + 0.4305\waterCh = 0.
\end{equation}

We can now incorporate the effect of this change in our probabilistic macrochemical model in equations~(7) through (14). This will impact only the equations for the estimation of measured compounds involved in the anabolic reaction. In this example, this modifies only equations~(10) and (13), which become
\begin{align}
  \mbox{Bicarbonate:}\ &
    \Delta\bicarbonateRV \sim \NormalDistrib(
      -\Delta\lactateRV_c - \frac{0.0063}{0.3354}\Delta\lactateRV_a,
      2\sigma_{\bicarbonateRV}^2) \\
  \mbox{Biomass:}\ &
    \Delta\biomassRV = \frac{1}{-0.3354}\Delta\lactateRV_a
\end{align}
and modifies the correspond lines in the model code (cf.~\sectionref{sec:model.code}).

\begin{figure}
  \begin{center}
    \subfloat[Estimates assuming the same cell weight across conditions.]{
      \includegraphics[width=0.46\textwidth]{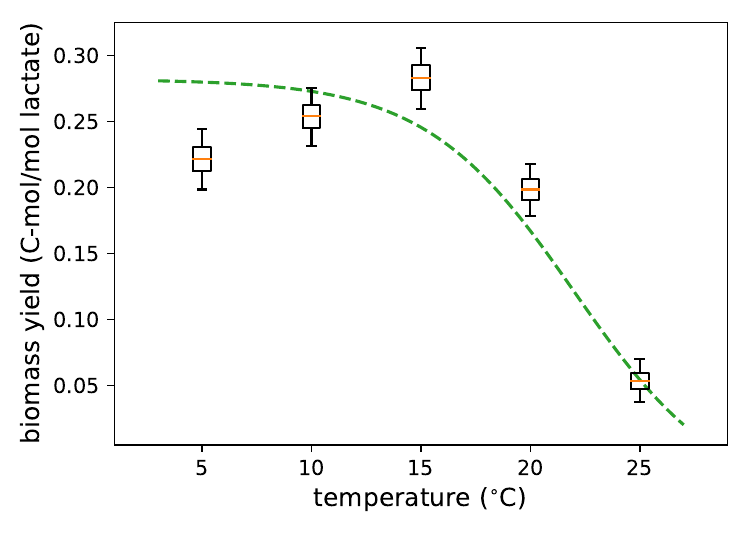}}
    \subfloat[Estimates with cell weight estimated per condition.]{
      \includegraphics[width=0.46\textwidth]{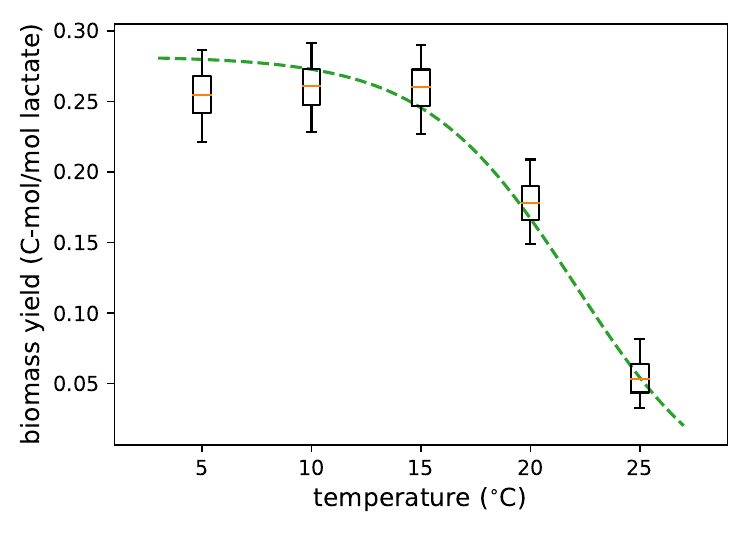}}
  \end{center}
  \caption{Biomass yield estimates using the probabilistic macrochemical model considering the modified biomass composition formula. Compare with Figures~6(b) and 6(c).}
  \label{fig:chem.model3:yield}
\end{figure}


The yield estimates considering this biomass composition formula are shown in \figref{fig:chem.model3:yield} for both the case of constant cell weight and cell weight per condition. Generally speaking, the new biomass yield estimates are slightly higher than before ($\le 3.7\%$). Much of that increase can be explained by the fact that the conversion factor between lactate consumed in the anabolic reaction to biomass produced increased $4.4\%$ from $2.857=1/0.35$ to $2.9815=1/0.3354$. The change is usually smaller because of the estimate of $\Delta\lactateRV_a$ also being smaller in this case. This can be understood by noting that the stoichiometric coefficient of bicarbonate in the anabolic reaction is almost an order of magnitude smaller in this case. Hence, the increased role of the total bicarbonate measurement in the catabolic reaction will tend to suggest a slightly larger estimate of $\Delta\lactateRV_c$ and, since the total lactate consumed is constant, this results in smaller estimates of $\Delta\lactateRV_a$.

\section{Macrochemical model implementation}
\label{sec:model.code}

The Python code for generating the data and Jupyter Python notebooks for replicating the results on the paper in their entirety are available at \url{github.com/arpaiva/biopgm-macrochem}.

For reference, the STAN probabilistic programming language implementation of the probabilistic macrochemical model, with cell weights estimated per condition, is also included herein. The variations used in each of the cases considered in the paper are shown in the Jupyter notebooks.
\lstset{basicstyle=\footnotesize, tabsize=2, numbers=left, numberstyle=\tiny}
\listCode{chemical_model_2b_cellWeightPerCond.stan}

\bibliographystyle{IEEEtran}
\bibliography{refs}